# The lower mass limit for circumbinary disc fragmentation

Matthew Teasdale ⋆ and Dimitris Stamatellos ⋆
*Jeremiah Horrocks Institute for Mathematics, Physics and Astronomy, University of Lancashire, Preston PR1 2HE, UK*

## ABSTRACT

In recent years, many wide orbit circumbinary giant planets have been discovered; some of these may have formed by gravitational fragmentation of circumbinary discs. The aim of this work is to investigate the lower mass limit for circumbinary disc fragmentation. We use the smoothed particle hydrodynamics code SEREN, which employs an approximate method for the radiative transfer, to perform three sets of simulations of gravitationally unstable discs. The first set of simulations covers circumstellar discs heated by a single 0.7 $M_{\odot}$ star (circumstellar model), the second set covers binaries with the same total stellar mass as the circumstellar model, attended by circumbinary discs with the same temperature profile (circumbinary fiducial model), and the third set covers circumbinary discs heated by each individual star (circumbinary realistic model). We vary the binary separation, mass ratio, and eccentricity to see their effect on disc fragmentation. For the circumstellar disc model, we find a lower disc-to-star mass ratio for fragmentation of $\sim 0.31$. For the circumbinary fiducial disc model we find the same disc-to-star mass ratio for fragmentation (but slightly lower for more eccentric, equal-mass binaries; 0.26). On the other hand, realistic circumbinary discs fragment at a lower mass limit (disc-to-star mass ratio of 0.17–0.26), depending on the binary properties. We conclude that circumbinary discs fragment at a lower disc mass (by ~45 per cent) than circumstellar discs. Therefore, gas giant planet around binaries may be able to form by gravitational instability easier than around single stars.

## 1 INTRODUCTION

Observations of young, protostellar objects have revealed discs around them. These discs show substructure such as gaps and rings, as well as spiral arms (S. M. Andrews et al. 2018). Spiral structures observed in protostellar discs, such as Elias 2–27, have been attributed to gravitational instability (F. Meru et al. 2017; C. Hall et al. 2018). Another example is the disc around AB Aurigae, which exhibits global spirals that are attributed to gravitational instability, as evidenced by kinematic signatures (J. Cadman, K. Rice & C. Hall 2021; J. Speedie et al. 2024). Using the gravitational instability 'wiggle' (C. Hall et al. 2020; C. Longarini et al. 2021), J. Speedie et al. (2024) estimate a total disc mass that is about 1/3 of the mass of the hosting star. In the same system a possible disc-instability Jovian planet has been observed at a wide separation (~93 au; T. Currie et al. 2022, 2025). In general, spiral arms that form as a result of gravitational instability transport angular momentum outwards in the disc, allowing for accretion on to the central star(s) (D. Lynden-Bell & A. J. Kalnajs 1972; C. Longarini et al. 2024).

Protostellar discs have been observed around binaries, with two examples of resolved circumbinary discs being GG Tau (S. Guilloteau, A. Dutrey & M. Simon 1999) and HD 142 527 (M. Fukagawa et al. 2006; A. P. Verhoeff et al. 2011). Observed circumbinary discs show features such as cavities and spirals caused by binary–disc interactions (M. Keppler et al. 2020; S. Hunziker et al. 2021; A. B. T. Penzlin et al. 2024). These observations have been complemented by simulation-based studies that look into multiple circumbinary disc and stellar parameters (M. M. Mutter, A. Pierens & R. P. Nelson 2017; J. Calcino et al. 2019; K. Hirsh et al. 2020; A. B. T. Penzlin et al. 2022; M. Teasdale & D. Stamatellos 2023).

Since the discovery of Kepler-16b (L. R. Doyle et al. 2011), $\sim 50$ circumbinary exoplanets have been confirmed, several of which may form in circumbinary protostellar discs (P. Delorme et al. 2013; B. Quarles et al. 2018; A. B. T. Penzlin et al. 2024; M. Teasdale & D. Stamatellos 2024). A circumbinary, or P-type planet, is defined as a planet that orbits a binary star. The properties of the binaries that host these exoplanets vary in separation, mass ratio and eccentricity. Fig. 1 shows the observed binary separations, $\alpha_{\rm b}$, plotted against the binary mass ratio, $q_{\rm b}$, and eccentricity, $e_{\rm b}$, for the currently known systems that host circumbinary exoplanets (with data taken from the NASA Exoplanet Archive and the Exoplanet.eu data base). The host binaries show a variety in separations ($10^{-3}$–20 au), mass ratios (0.05–1), and eccentricities (0–0.9), which suggests a variety in the properties of the exoplanets that they host. The number of observed circumbinary exoplanets is limited but it is likely that many wide-orbit P-type planets are yet to be discovered.

It has been argued that the formation of gas giant planets, especially those on wide orbits, may be possible through gravitational instability (G. P. Kuiper 1951; A. G. W. Cameron 1978; A. P. Boss 1997; D. Stamatellos & A. P. Whitworth 2009a; A. Fenton & D. Stamatellos 2024). Planetary systems like HR 8799 (C. Marois et al. 2008, 2010), or the Jovian planet around AB Aurigae (T. Currie et al. 2022, 2025) cannot have formed through core accretion as this mechanism in inefficient at large distances from the central star. Gravitational instability may also be be possible to form even lower mass (super-Earth) planets either through tidal downsizing (S. Nayakshin 2017) or through facilitating dust growth and fragmentation of the dust component (W. K. M. Rice et al. 2006; C. Longarini et al. 2023a,b). It has also been suggested that the inclusion of magnetic fields may reduce the mass of fragments formed by gravitational instability, leading to the formation of intermediate-mass planets (H. Deng, L. Mayer & R. Helled 2021).

⋆ E-mail: MTeasdale1@lancashire.ac.uk (MT); DStamatellos@lancashire.ac.uk (DS)

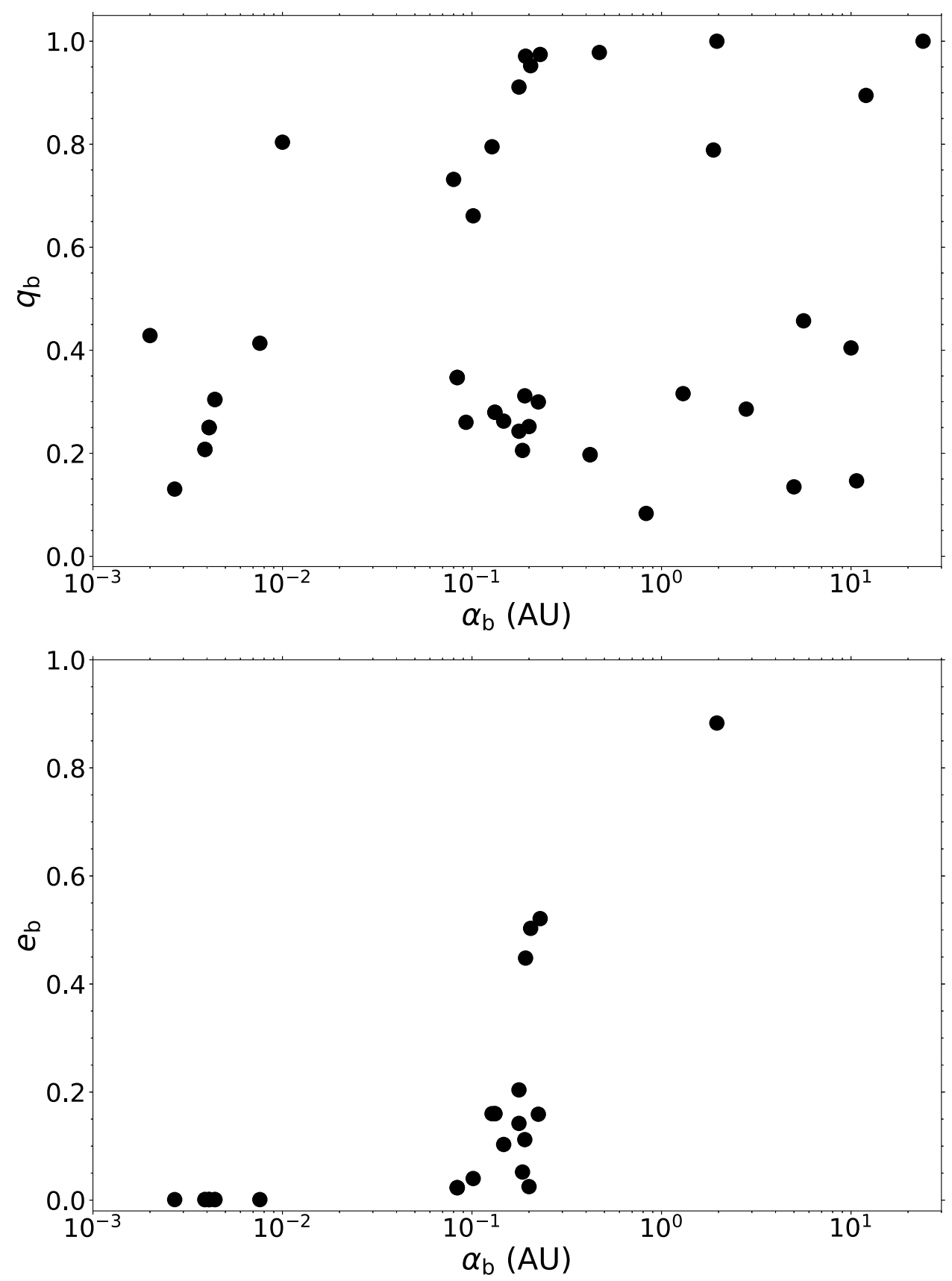

**Figure 1.** The separation of the host binaries of the currently known systems with circumbinary planets, plotted against their mass ratio (top) and eccentricity (bottom). Data from the 'Extrasolar Planets Encyclopaedia' (https://exoplanet.eu/) and from the 'NASA Exoplanet Archive (https://exoplanetarchive.ipac.caltech.edu) which is operated by the California Institute of Technology, under contract with the National Aeronautics and Space Administration under the Exoplanet Exploration Program (J. L. Christiansen et al. 2025).

A protostellar disc becomes gravitationally unstable if the Toomre criterion is satisfied (A. Toomre 1964),

$$Q = \frac{c_s \Omega}{\pi G \Sigma} \lesssim Q_{\rm crit} \simeq 1 - 2, \tag{1}$$

where $Q$ is the Toomre parameter, $c_s$ is the sound speed, $\Omega$ the angular frequency, $G$ the gravitational constant, and $\Sigma$ the surface density of the disc. Effectively $Q$ measures the balance between the thermal and rotational support of the disc against gravity. If a disc is sufficiently massive, so that $Q \simeq 1$, a spiral structure forms. For the gravitational instability to lead to fragmentation the disc needs to cool on a short time-scale (i.e. $\tau_c \lesssim 3\,\Omega^{-1}$) (C. F. Gammie 2001; B. M. Johnson & C. F. Gammie 2003; W. K. M. Rice et al. 2003; W. K. M. Rice, G. Lodato & P. J. Armitage 2005). Circumbinary planets may also form the same way; an example of a circumbinary disc that undergoes fragmentation is L1448 IRS3B (J. J. Tobin et al. 2016; N. K. Reynolds et al. 2021).

It is important to investigate the conditions under which circumbinary discs may fragment, in order to determine whether circumbinary planets can form through this process. The physics of fragmentation of circumstellar discs has been explored in great detail (e.g. W. K. M. Rice et al. 2003; D. Stamatellos & A. P. Whitworth 2009a; D. Stamatellos et al. 2011; F. Meru & M. R. Bate 2012; H. Deng, L. Mayer & F. Meru 2017; C. Hall, D. Forgan & K. Rice 2017; J. Cadman et al. 2020). While there is a dependence on the specific disc and stellar parameters, simulations that include detailed radiative heating and cooling, find that fragmentation is possible for a minimum disc-to-star mass ratio of $q_{\rm D} \sim 0.3$ (D. Stamatellos et al. 2011; T. J. Haworth et al. 2020; A. Mercer & D. Stamatellos 2020). D. Stamatellos et al. (2011) studied circumstellar discs around a 0.7 $M_{\odot}$ star and found that fragmentation can happen for disc-to-stellar mass ratios of at least 0.36. A. Mercer & D. Stamatellos (2020) studied M dwarfs (0.2–0.4 $M_{\odot}$) and found a disc-to-stellar mass ratio from at least 0.3 to 0.6 for fragmentation, with a dependence on the stellar mass and disc radius. They find that the lower mass limit for disc fragmentation increases linearly with the stellar mass and that larger discs require higher mass for fragmentation to happen. A. Mercer & D. Stamatellos (2020) also find that higher disc-to-stellar mass ratios are needed for fragmentation of discs around lower mass stars. This result is corroborated by J. Cadman et al. (2020) who studied disc around higher-mass stars (0.25–2 $M_{\odot}$). This means that lower-mass stars may be able to support discs with higher disc-to-stellar mass ratios, without fragmentation happening (T. J. Haworth et al. 2020).

The aim of this work is to investigate the lower mass limit for circumbinary disc fragmentation and compare it with the limit for circumstellar discs. We describe the computational method in Section 2, and in Section 3 the simulation set-up. In Section 4, we present the results on the minimum disc mass required for fragmentation, and in Section 5 we make comparisons between circumstellar and circumbinary discs. We finally discuss the wider implications of this work and its conclusions in Section 6.

## 2 COMPUTATIONAL METHOD

We use SEREN, a three-dimensional smooth particle hydrodynamics (SPH) code developed by D. A. Hubber et al. (2011). The code includes the effects of radiative transfer using the approximate method of J. C. Lombardi, W. G. McInally & J. A. Faber (2015). This method determines the heating and cooling rate of the gas within the protostellar disc, as

$$\frac{{\rm d}u_{\rm i}}{{\rm d}t} = \frac{4\sigma_{\rm SB}(T_{\rm A}^4 - T_{\rm i}^4)}{\bar{\Sigma}_{\rm i}^2 \bar{\kappa}_{\rm R}(\rho_{\rm i}, T_{\rm i}) + \kappa_{\rm P}^{-1}(\rho_{\rm i}, T_{\rm i})}, \tag{2}$$

where $u_{\rm i}$ is the specific internal energy of the SPH particle, $\sigma_{\rm SB}$ is the Stefan–Boltzmann constant, $T_{\rm A}$ the pseudo-background temperature above which gas can cool radiatively ($T_{\rm A}$ is due to heating from the stars in the system), $\bar{\Sigma}$ the mass-weighted mean column density, $\rho_{\rm i}$ is the density of the particle, $T_{\rm i}$ is the temperature of the particle, $\bar{\kappa}_{\rm R}$ and $\bar{\kappa}_{\rm P}$ are the Rosseland and Planck-mean opacities. In our case, $T_{\rm A}$ is the background temperature set by the stars of the system. The mass-weighted mean column density, which regulates heating and cooling is calculated using the pressure scale-height which is a good

**Table 1.** The stellar parameters used for the CBR model simulations. $q_b$ is the binary mass ratio, $M_\star$ is the mass of each star of the binary, and $T_i$ is the temperature at 1 au from each star of the binary.

| $q_b$ | $M_\star$ ($M_\odot$) | $T_i$ (K) |
|---|---|---|
| 1 | 0.35 | 130 |
| | 0.35 | 130 |
| 0.3 | 0.53 | 180 |
| | 0.16 | 85 |
| 0.1 | 0.64 | 220 |
| | 0.064 | 40 |

approximation in discs,

$$\bar{\Sigma} = \zeta' \frac{P}{|\boldsymbol{a}_h|}, \quad (3)$$

where $\zeta' = 1.014$ is a dimensionless coefficient, $P$ the gas pressure, and $\boldsymbol{a}_h$ the hydrodynamical acceleration (i.e. without including the gravitational or viscous contribution),

$$\boldsymbol{a}_h = \frac{-\nabla P}{\rho}. \quad (4)$$

This method is used as it is a better approximation of the radiative transfer processes in protostellar discs than the D. Stamatellos et al. (2007) method that uses the gravitational potential as a proxy for the column density (A. Mercer, D. Stamatellos & A. Dunhill 2018; A. K. Young et al. 2024). This method results in more efficient cooling than the D. Stamatellos et al. (2007) approximation, with the expectation that discs may fragment at lower masses.

## 3 SIMULATION SET-UP

We perform three sets of simulations. The first set of simulations covers circumstellar discs heated by a single star (referred to as the *circumstellar* model; CS) as e.g. in A. Mercer & D. Stamatellos (2020), the second set covers circumbinary discs with the same temperature profile as the circumstellar disc model (referred to as the *fiducial* model; CBF), and the third set covers circumbinary discs that are heated by each star individually (referred to as the *realistic* model; CBR). Our goal is to compare fragmentation of circumbinary discs with that of circumstellar discs, and to investigate how the initial binary parameters (i.e. binary mass ratio, separation, eccentricity, and temperature profile) affects fragmentation.

We use a stellar mass of $M_\star = 0.7\,M_\odot$ for the CS model. For the circumbinary disc models (CBF, CBR) the total mass of the binary is set to $M_\star = 0.7\,M_\odot$, with individual stellar masses detailed in Table 1. We model three initial binary mass ratios, $q_B = 1$, 0.3, and 0.1, two initial binary separations, $\alpha_b = 5$, 10 au, and two initial binary eccentricities, $e_b = 0.2$ and 0.5. These values were chosen to reflect the typical properties of binaries known to host circumbinary exoplanets (see Fig. 1). We do not model binaries on circular orbits as test simulations showed that these binaries quickly become eccentric (R. M. Heath & C. J. Nixon 2020; D. Lai & D. J. Muñoz 2023; M. Siwek, R. Weinberger & L. Hernquist 2023), especially since the disc mass is rather high. The stars in the simulations are represented with sink particles with accretion radius $R_s^\star = 1$ au.

For all models, we assume a disc extending from $R_{in}^D = 5$ au to $R_{out}^D = 120$ au that is represented by $10^6$ SPH particles (as e.g. in A. Mercer & D. Stamatellos 2020). This number ensures that gravitational fragmentation is properly resolved. M. R. Bate & A. Burkert (1997) argue that the Jeans mass must be resolved by $2 \times N_{neigh}$ and A. F. Nelson (2006) conclude that the Toomre mass must be resolved by $6 \times N_{neigh}$, where $N_{neigh} \approx 50$ is the number of SPH neighbours. The simulations performed by D. Stamatellos & A. P. Whitworth (2009b) with a 0.7 $M_\odot$ disc and $1.5 \times 10^5$ particles find a minimum Jeans mass of $\sim 2\,M_J$ and a minimum Toomre mass of $\sim 2.5\,M_J$. If we take $2\,M_J$ as a lower resolution limit, then this corresponds to $\sim 128 \times N_{neigh}$ i.e. their disc is sufficiently resolved. In this study we use a larger number of SPH particles and a lower mass disc, therefore the above conditions are comfortably met. The vertical structure of our discs is also adequately resolved, since we use ~7 times more particles than in their simulations, in which the disc scale height is resolved by a factor of more than 3–5 smoothing lengths.

The surface density profile of the disc is set to

$$\Sigma(R) = \Sigma(1\,\mathrm{AU}) \left(\frac{R}{\mathrm{AU}}\right)^{-1}, \quad (5)$$

for all three different models.

For the CS model the disc temperature is set to

$$T(R) = 250\mathrm{K} \left(\frac{R}{\mathrm{AU}}\right)^{-0.7} + 10\,\mathrm{K}, \quad (6)$$

where $\Sigma$(1 AU) is determined by the radius and mass of the disc, and $R$ is the distance to the central star. We chose a temperature power index of 0.7 as observations find typical values from 0.35 to 0.8 (S. M. Andrews et al. 2009). For the circumbinary fiducial model, we use the same temperature profile for the disc as above, with $R$ being the distance to the centre of mass of the binary. For the circumbinary realistic model, we include heating from both stars, setting the disc temperature profile to

$$T(R_1, R_2) = \sum_{i=1}^{2} \left\{ T_i(1\,\mathrm{AU}) \left(\frac{R_i}{\mathrm{AU}}\right)^{-0.7} \right\} + 10\,\mathrm{K}, \quad (7)$$

where $T_i$(1AU) is the temperature at 1 au from the corresponding star (see Table 1), when ignoring the other star, $R_1$ the distance from the primary, and $R_2$ the distance from the secondary star of the binary, respectively. In effect, we assume that at any given location in the disc the temperature is due to implicit heating from both stars; the temperature is the sum of the contributions from each star. Our model therefore includes asymmetric radiation from the binary components, which has important implications on the disc dynamics (P. P. Poblete et al. 2025). The temperatures in Table 1 were selected using typical luminosities for each stellar mass (e.g. C. Cifuentes et al. 2020), and assuming that the temperature scales as $T(R) = (L_\star/16\pi\sigma_{SB})R^{-1/2}$. The initial disc temperature profiles, as given in equations (6)–(7), also act as the pseudo-background temperature $T_A$ for the approximate radiative transfer method (see equation 2).

## 4 THE MINIMUM MASS LIMIT FOR FRAGMENTATION OF CIRCUMBINARY DISCS

We first simulate discs with mass $M_D = 0.3\,M_\odot$ for each parameter configuration as this mass leads to $Q \sim 0.5$, i.e. highly unstable discs. In a typical simulation spiral arms form that eventually fragment to form planets and higher mass objects (brown dwarfs and low mass stars), in agreement with previous simulations (e.g. D. Stamatellos & A. P. Whitworth 2009a; A. Mercer & D. Stamatellos 2017). We assume that fragmentation is achieved when a condensation reaches a density of $10^{-9}$ g cm$^{-3}$, whereby a sink with an accretion radius of $R_s = 0.1$ au is introduced to represent the fragment. At this density the first hydrostatic core has been securely formed (in fact this happens at around $10^{-12}$ g cm$^{-3}$; D. Stamatellos & A. P. Whitworth 2009b; A. Mercer & D. Stamatellos 2020; A. Fenton & D. Stamatellos 2024) and the clump is gravitationally bound. To determine the lower limit for disc fragmentation we progressively reduce the disc mass by 0.02 $M_\odot$, and repeat each simulation until the disc does not fragment. We perform three sets of simulations (CS, CBF, CBR) and for each combination of binary parameters we perform three simulations with different realization of the disc, to account for the stochastic nature of disc fragmentation. The lower limit for fragmentation is identified as the disc mass at which even one of the three simulated discs fragments. The results on the lower mass limit for disc fragmentation are summarized in Table 2. Each row in this table corresponds to three simulations with the same initial conditions (but with different realization of the disc, i.e. different random seeds); i.e. there is a total of 75 simulations.

**Table 2.** The lower mass limit for disc fragmentation for the fiducial, realistic, and circumstellar models. Here, Type refers the CS, CBF, CBR models, $\alpha_b$ is the initial binary separation, $q_b$ the binary mass ratio, $e_b$ the binary eccentricity, $M_D$ the lower mass limit for disc fragmentation and $q_D$ the lower disc-to-star/binary mass ratio for fragmentation.

| Type | $\alpha_b$ (au) | $q_b$ | $e_b$ | $M_D$ ($M_\odot$) | $q_D$ |
|---|---|---|---|---|---|
| CS | | | | 0.22 | 0.31 |
| CBF | 10 | 1 | 0.2 | 0.20 | 0.29 |
| | | | 0.5 | 0.18 | 0.26 |
| | | 0.3 | 0.2 | 0.22 | 0.31 |
| | | | 0.5 | 0.22 | 0.31 |
| | | 0.1 | 0.2 | 0.22 | 0.31 |
| | | | 0.5 | 0.22 | 0.31 |
| | 5 | 1 | 0.2 | 0.22 | 0.31 |
| | | | 0.5 | 0.22 | 0.31 |
| | | 0.3 | 0.2 | 0.22 | 0.31 |
| | | | 0.5 | 0.22 | 0.31 |
| | | 0.1 | 0.2 | 0.22 | 0.31 |
| | | | 0.5 | 0.22 | 0.31 |
| CBR | 10 | 1 | 0.2 | 0.16 | 0.23 |
| | | | 0.5 | 0.16 | 0.23 |
| | | 0.3 | 0.2 | 0.16 | 0.23 |
| | | | 0.5 | 0.14 | 0.20 |
| | | 0.1 | 0.2 | 0.16 | 0.23 |
| | | | 0.5 | 0.16 | 0.23 |
| | 5 | 1 | 0.2 | 0.18 | 0.26 |
| | | | 0.5 | 0.12 | 0.17 |
| | | 0.3 | 0.2 | 0.18 | 0.26 |
| | | | 0.5 | 0.12 | 0.17 |
| | | 0.1 | 0.2 | 0.18 | 0.26 |
| | | | 0.5 | 0.12 | 0.17 |

### 4.1 Circumstellar model

The evolution of the surface density of a representative circumstellar disc, at the disc fragmentation limit, is shown in Fig. 2. We find the lower mass limit for disc fragmentation with our parameters is $M_D = 0.22\,M_\odot$, which corresponds to a disc-to-star mass ratio of $q_D = 0.31$ (see Table 2). Our result is in agreement with T. J. Haworth et al. (2020) who use the same stellar mass and find that fragmentation occurs down to $q_D \sim 0.3$, using a disc of radius 100 au and the radiative transfer approximation of D. Forgan et al. (2009).

This value is lower than that found by D. Stamatellos et al. (2011) ($q_D = 0.36$) despite using the same system parameters. We credit this to the use of the J. C. Lombardi et al. (2015) radiative transfer approximation method as opposed to the D. Stamatellos et al. (2007) method; the former results in more efficiently disc cooling.

J. Cadman et al. (2020) find that fragmentation occurs for a disc-to-star mass ratio of $q_D = 0.5$, using a solar mass star and a disc radius of 140 au. Their estimated $q_D$ is expectedly higher than what we find due to simulating a larger radii disc than we do, and the use of the D. Stamatellos et al. (2007) method for the radiative transfer. Both factors limit the likelihood of fragmentation due to lower disc surface density (as a result of the larger disc radius) and the disc not being able to cool efficiently enough.

A. Mercer & D. Stamatellos (2020) use the same radiative transfer approximation but a different SPH code (GANDALF; D. A. Hubber, G. P. Rosotti & R. A. Booth 2018) and for a disc with radius 120 au they find that the minimum disc mass for fragmentation is $M_{\rm disc}^{\rm 120AU} = 0.08 M_\odot + 0.22(M_\star/M_\odot)$. Substituting for a stellar mass of 0.7 $M_\odot$ that we use, we estimate a minimum disc mass of 0.23 $M_\odot$, i.e. a disc-to-star mass ratio of $q_D = 0.33$, which is slightly higher than what we find here. However, the previous relation was derived from simulations of M dwarfs with masses up to 0.4 $M_\odot$, i.e. a lower stellar mass than in our case.

### 4.2 Circumbinary fiducial model

Figs 3 and 4 show snapshots of the disc surface density for representative circumbinary fiducial simulations, at a time just before fragmentation occurs, for all the binary parameters investigated (see Table 2). The majority of the simulations presented here are at the lower mass limit for disc fragmentation.

In the simulations with a lower binary mass ratio (i.e. $q_b = 0.3, 0.1$), we find that the lower mass limit for fragmentation is the same as in the circumstellar disc model, i.e. at a disc mass of $M_D = 0.22\,M_\odot$ (see Table 2). This is to be expected as (i) the stellar heating is exactly the same as in the CS model (per construction), and (ii) due to small binary separation and mass ratio, the two stars have the same gravitational effect on the disc as a single star with the same combined mass.

On the other hand, for the high binary mass ratio, $q_b = 1$, and larger binary separation, $\alpha_b = 10$ au, fragmentation is possible at a lower disc mass. For a binary eccentricity of $e_b = 0.2$, fragmentation is possible at $M_D = 0.2\,M_\odot$, whereas for $e_b = 0.5$, fragmentation is possible at $M_D = 0.18\,M_\odot$ with the corresponding disc-to-star mass ratios $q_D = 0.26$ and 0.29, respectively. This lower mass limit is expected as the gravitational effect of the binary on the disc is more pronounced when the binary 'deviates' from a single star, i.e. the binary separation, mass ratio and the eccentricity are high (D. Lai & D. J. Muñoz 2023; M. Teasdale & D. Stamatellos 2023). This induces structure in the disc that promotes fragmentation at slightly lower disc masses than in the CS model, despite the fact that the disc temperature profiles are almost the same.

### 4.3 Circumbinary realistic model

This model includes asymmetric radiation from the binary components, which affects the disc dynamics (P. P. Poblete et al. 2025). Snapshots for representative circumbinary realistic simulations near the fragmentation limit are shown in Figs 5 and 6.

We find that the mass needed for the realistic circumbinary disc to fragment is lower than that of the fiducial CB model by $\sim 45$ per cent, with a minimum disc mass of $M_D = 0.12\,M_\odot$ and disc-to-star mass ratio of $q_D = 0.17$ (see Table 2). This is because in the realistic model the heating provided by the stars is lower than that provided in the fiducial model; the stellar temperatures used for $T_{\rm i}$ [see equation (7) and Table 1] are lower than the 250 K used in the circumstellar and circumbinary fiducial models, which lead to a cooler disc that, for the same mass, has a lower Toomre parameter Q. Additionally, due to their lower temperature these discs cool more efficiently (see equation 2).

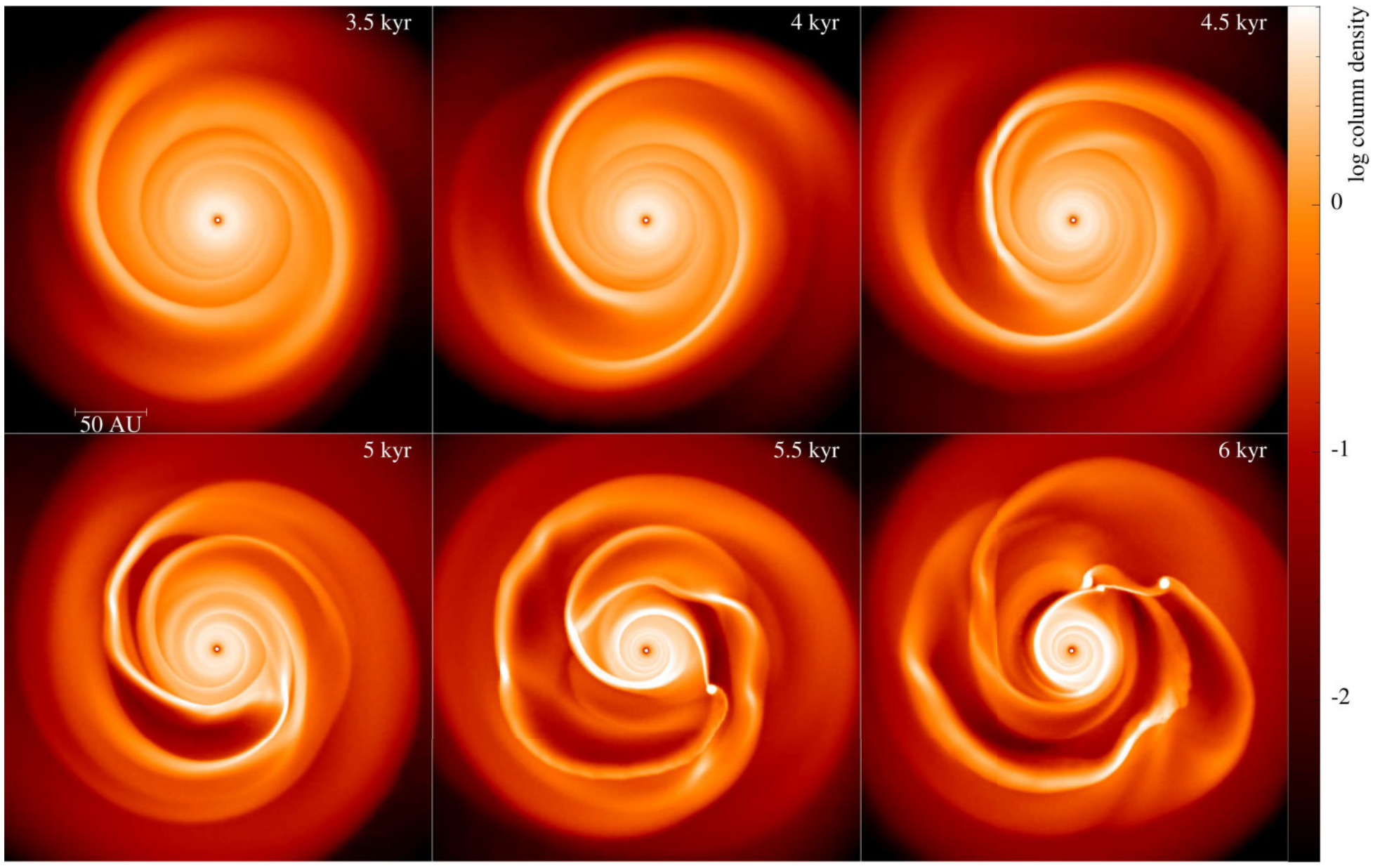

**Figure 2.** The evolution of the disc surface density (g cm$^{-2}$) for a circumstellar disc of mass $M_{\rm D} = 0.22\,{\rm M}_{\odot}$, with stellar and disc parameters discussed in Section 3.

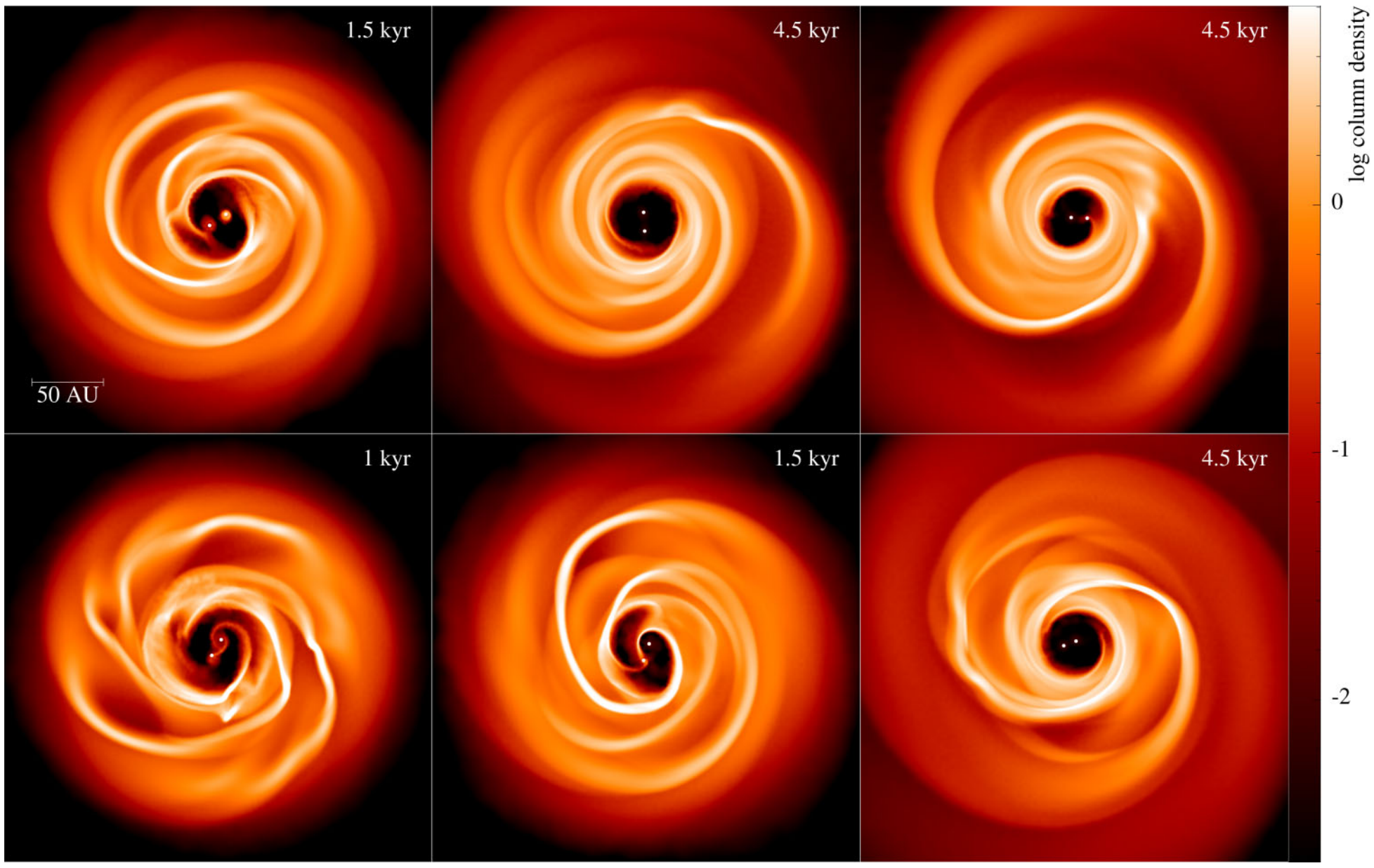

**Figure 3.** Disc surface density (g cm$^{-2}$) snapshots for representative circumbinary fiducial model simulations with a disc mass of $M_{\rm D} = 0.22\,{\rm M}_{\odot}$, and a binary separation of $\alpha_{\rm b} = 10$ au. The first column corresponds to simulations with a binary mass ratio of $q_{\rm b} = 1$, the second with a binary mass ratio of $q_{\rm b} = 0.3$, and the third with a binary mass ratio of $q_{\rm b} = 0.1$. The top row corresponds to simulations with a binary eccentricity of $e_{\rm b} = 0.2$ and the bottom row with a binary eccentricity of $e_{\rm b} = 0.5$.

In realistic circumbinary discs, there are two competing factors that affect the likelihood of fragmentation: (i) the gravitational effect of the binary on the disc, and (ii) disc heating from the components of the binary. Both of these are larger for high mass ratio, separation and eccentricity; however, the former promotes fragmentation by creating density enhancements in the disc, whereas the latter suppresses fragmentation by increasing the disc temperature. These two factors seem to cancel out in the case of wide binary simulations ($\alpha_{\rm b} = 10$ au) as the lower disc mass limit for fragmentation is at $0.16\,{\rm M}_{\odot}$, irrespective of the binary eccentricity and stellar mass ratio (apart from one case). On the other hand, in the close binary simulations ($\alpha_{\rm b} = 5$ au), discs around more eccentric binaries ($e_{\rm b} = 0.5$) fragment at lower mass than discs around less eccentric binaries ($e_{\rm b} = 0.2$), with the mass limit for fragmentation being 0.12 and $0.18\,{\rm M}_{\odot}$, respectively. We find no dependence

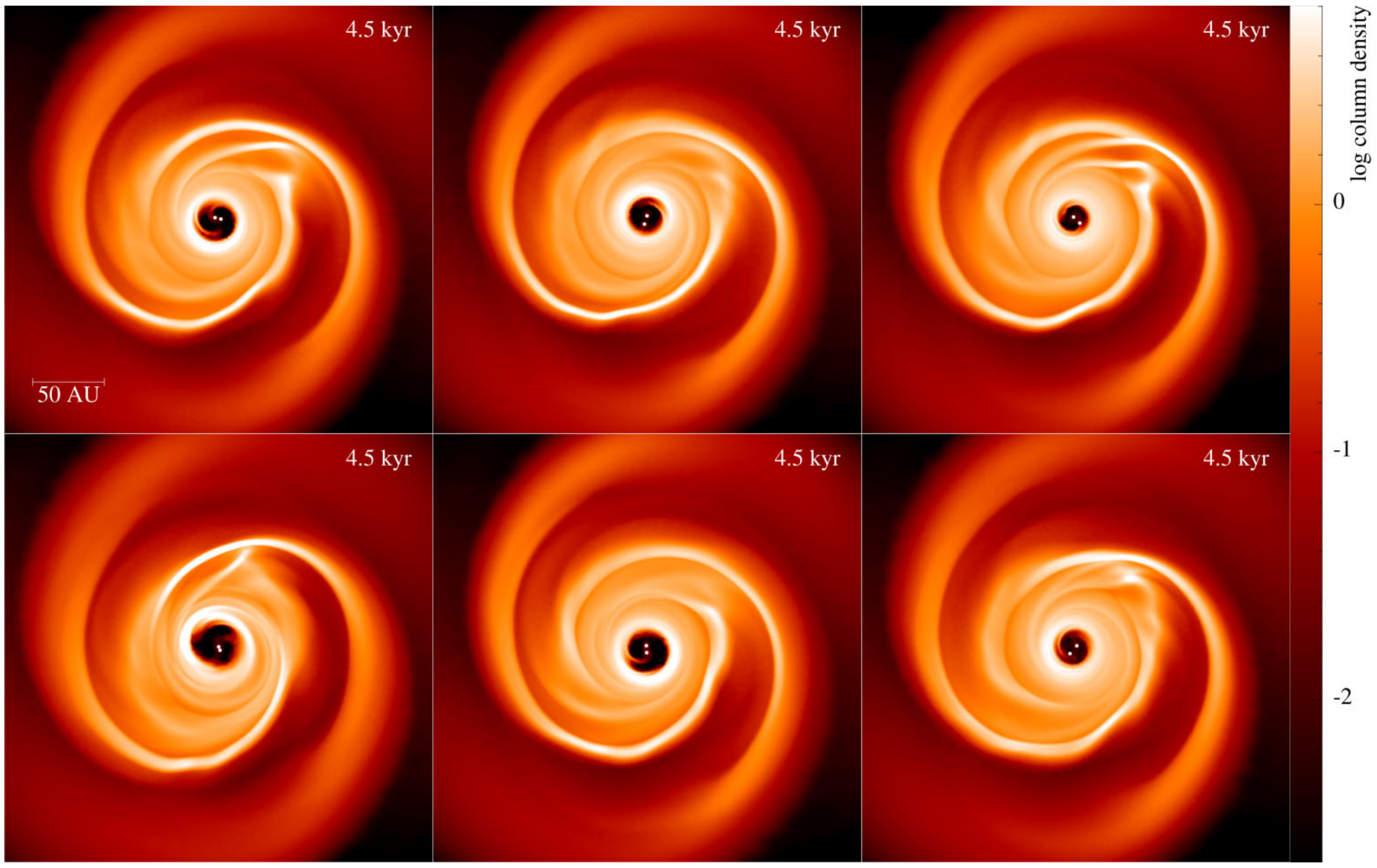

**Figure 4.** Same as in Fig. 3 but for CBF simulations with an initially binary separation of $\alpha_{\rm b} = 5$ au.

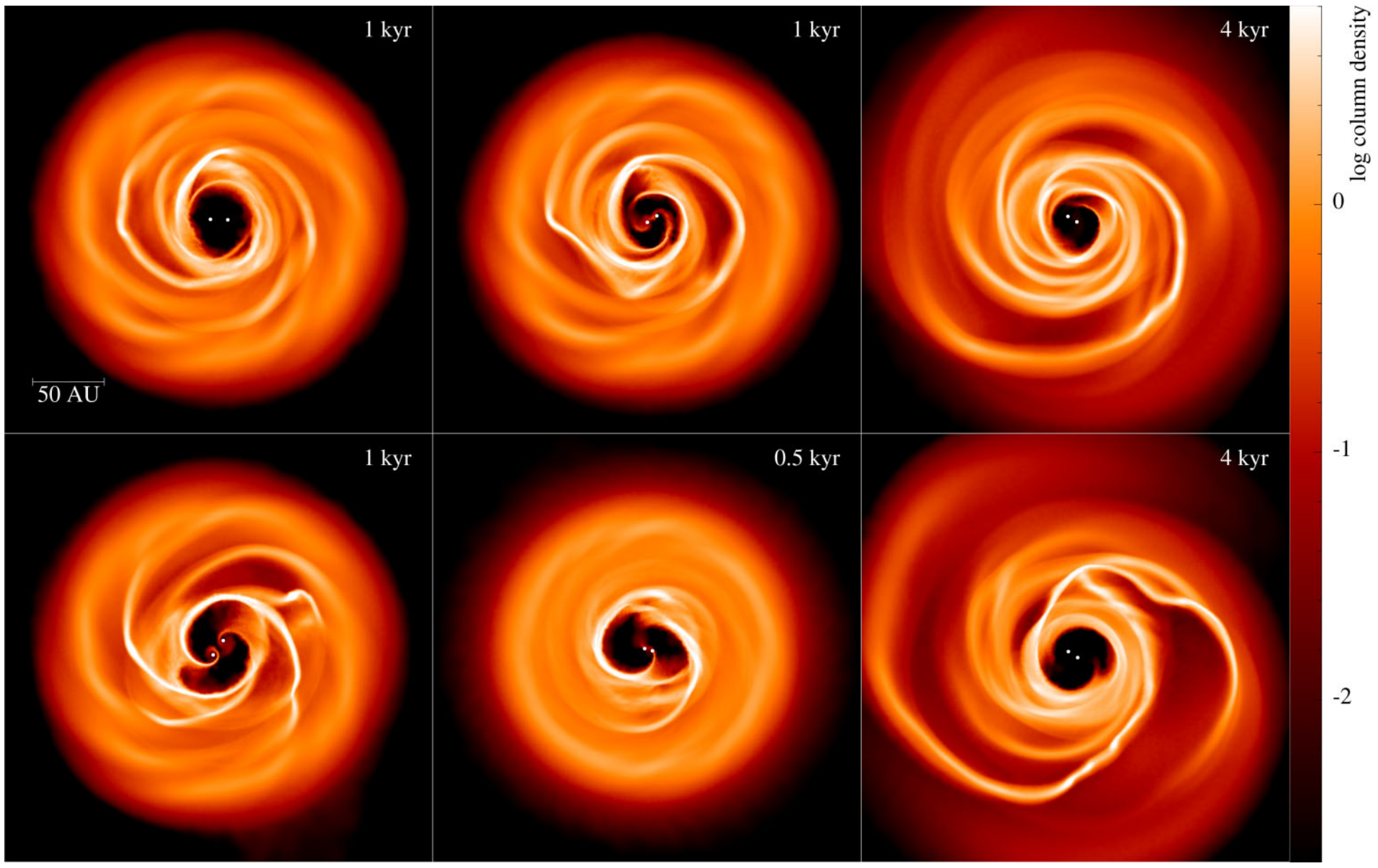

**Figure 5.** Disc surface density (g cm$^{-2}$) snapshots for representative circumbinary realistic model simulations with a disc mass of $M_{\rm D} = 0.18\,{\rm M}_{\odot}$, and a binary separation of $\alpha_{\rm b} = 10$ au. The first column corresponds to simulations with a binary mass ratio of $q_{\rm b} = 1$, the second with a binary mass ratio of $q_{\rm b} = 0.3$, and the third with a binary mass ratio of $q_{\rm b} = 0.1$. The top row shows simulations with a binary eccentricity of $e_{\rm b} = 0.2$ and the bottom row with a binary eccentricity of $e_{\rm b} = 0.5$.

of the lower mass limit for fragmentation on the binary mass ratio.

## 5 COMPARISON BETWEEN CIRCUMSTELLAR AND CIRCUMBINARY DISCS

We compare the disc morphology in the three sets of simulations (circumstellar, circumbinary fiducial, and circumbinary realistic models), using the Toomre parameter (with references to the temperature and surface density of the disc) and the strength of the spiral arms present in the disc, just before fragmentation. This comparison provides insights on the physics of disc fragmentation.

### 5.1 Disc morphology

The Toomre parameter, surface density and temperature profiles for representative circumstellar, circumbinary fiducial, and realistic models are shown in Figs 7 and 8. We use simulations with disc mass $M_{\rm D} = 0.22\,{\rm M}_{\odot}$ for the circumstellar and fiducial models, as this corresponds to the lower fragmentation mass (for the majority of the simulations). For the realistic model, we use the simulation with disc mass $M_{\rm D} = 0.18\,{\rm M}_\odot$ (chosen for the same reason). For all simulations, we take the snapshot just before fragmentation (at most 500 yr before fragmentation) to make comparisons. The surface density plots of the snapshots are shown in Figs 3–6.

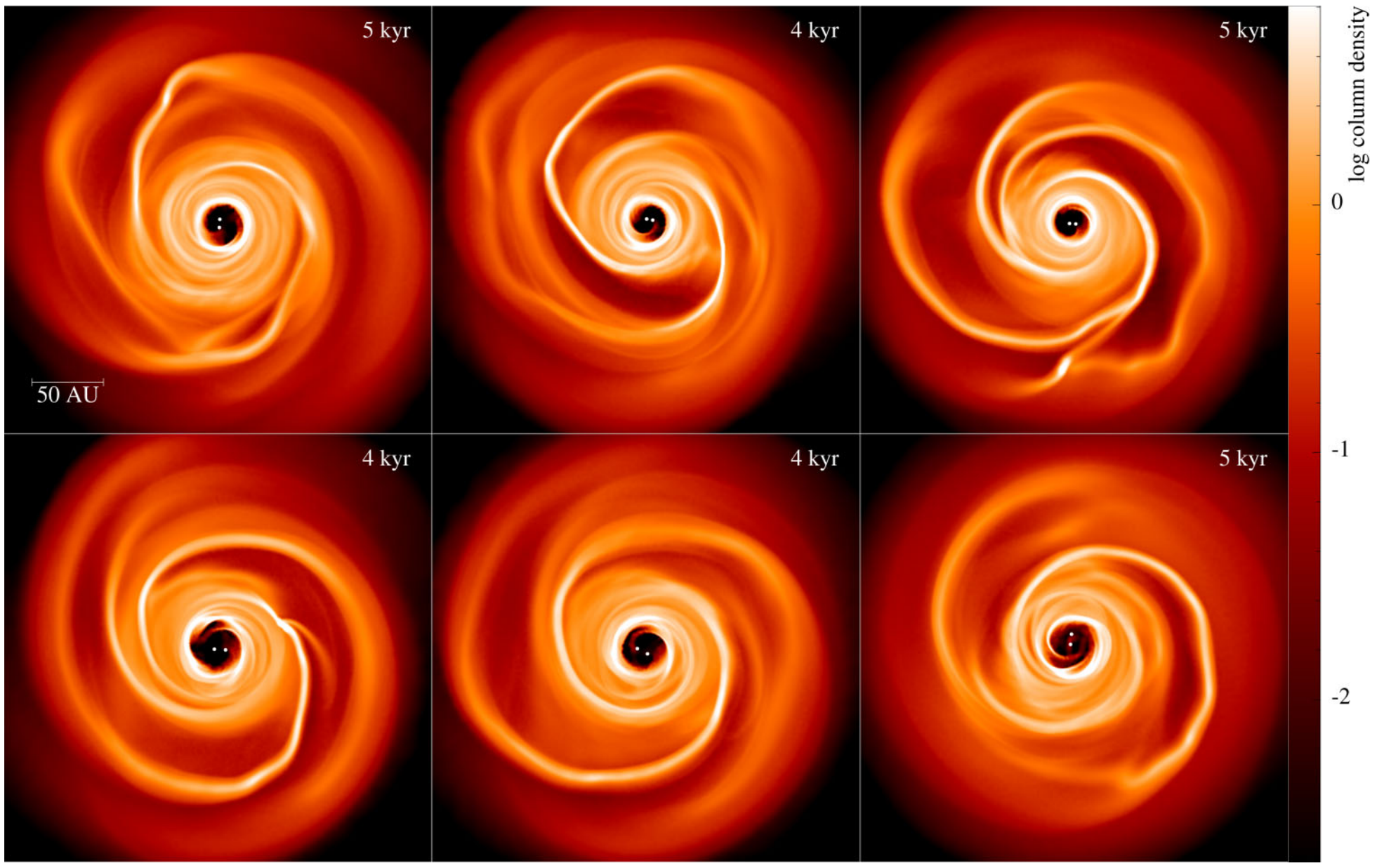

**Figure 6.** Same as in Fig. 5 but for CBR simulations with an initially binary separation of $\alpha_{\rm b} = 5$ au.

In the circumstellar model (Fig. 7), the Toomre criterion is satisfied ($Q \leq 1$) outside $\sim 40$ au. Furthermore, the temperature and surface density of the disc drops with distance from the star, albeit with several peaks ($\sim 15$, 45, and 85 au), which correspond to dense, hot condensations with the disc.

The fiducial model (Fig. 7) employs the same heating as the circumstellar model, and the disc shows similar behaviour. However, the density peaks are at different radii, with the $Q$ close to 1 even at a distance 20 au from the centre of mass of the binary.

In the circumbinary realistic disc model (Fig. 8), the disc is gravitationally unstable beyond 20 au (i.e. $Q \lesssim 1$–2), similarly with the other two cases. However there is more structure in the disc, as seen by the dips in $Q$ (or equivalently by the peaks in surface density). This is due to the temperature of the disc being significantly lower as opposed to the previous models. CBR models with wide binaries ($\alpha_{\rm b} = 10$ au) and a high binary mass ratio ($q_{\rm b} = 1, 0.3$) tend to become unstable at larger radii, than models with close binaries ($\alpha_{\rm b} = 5$ au).

## 5.2 Strength of spiral arms

We quantify the strength of the spiral arms, by calculating their amplitude, in order to correlate it with the development of gravitational instability and disc fragmentation for the different models. We use the method of J. P. Sleath & P. Alexander (1996), in which the logarithmic spiral is described by

$$R = R_0 {\rm e}^{-m\phi/\zeta}, \quad (8)$$

where $\phi$ is the azimuthal angle of the SPH particle, $m$ is the mode of the perturbation, and $\zeta = -m/\tan\beta$ represents the pitch angle $\beta$ of the spiral (J. P. Sleath & P. Alexander 1996). The amplitude $F(\zeta, m)$ of a specific mode $m$ is

$$F(\zeta, m) = \int_{-\infty}^{\infty}\int_{-\pi}^{\pi}\sum_{j=1}^{N}\{\delta(u - \ln[R_j]\delta(\phi - \phi_j)\} \times {\rm e}^{-{\rm i}(\zeta u + m\phi)}{\rm d}u{\rm d}\phi = \frac{1}{N}\sum_{j=1}^{N}{\rm e}^{-{\rm i}(\zeta\,\ln[R_j] + m\phi_j)}, \quad (9)$$

where $(R_j, \phi_j)$ are the co-ordinates of particle $j$. Fig. 10 shows the amplitude of the $m = 2$ mode of representative snapshots from the CBF model (see Fig. 3 for the corresponding surface density plots). We focus on the $m = 2$ mode as this generally the dominant one (see Fig. 9 that demonstrates this for a representative simulation). Fig. 11 shows the peak amplitude values (calculated for each snapshot across all pitch angles) for all three models close to the time of fragmentation (see Figs 3 and 4 for the CBF model and Figs 5 and 6 for the CBR model) plotted against the pitch angle of the spiral.

The discs in the circumbinary fiducial runs all show higher peak amplitudes than those of the circumstellar disc runs. This is despite the temperature profile of the circumstellar model being the same to that of the circumbinary fiducial model (see Fig. 7). This shows the effect of the binary on promoting stronger spirals. However, fragmentation does not happen at a lower disc mass, as this is regulated by the disc cooling (C. F. Gammie 2001); indeed, as the disc temperature profile (which acts as the pseudo-background temperature, $T_{\rm A}$, below which gas cannot cool radiatively) is the same for both the CS and CBF models, cooling is also similar (see equation 2). Therefore, in these two models, fragmentation happens at a similar disc mass despite the fact that the spiral arms are stronger for a circumbinary disc than for a circumstellar disc.

The discs in the circumbinary realistic model also generally show higher peak amplitudes than the discs in the CS model, but there are a few cases where the amplitude is slightly lower. The amplitudes are also generally lower than the peak amplitudes of the CBF model discs. Despite this, realistic CB discs fragment at lower disc masses and mass ratios (down to $0.18\,{\rm M}_\odot$ and 0.17, respectively, depending on the binary properties), than CBF and circumstellar discs (0.22 $M_\odot$ and 0.31). The reason for this is that due to the lower disc temperature in the CBR model [see equation (7) and Table 1] the disc can cool more efficiently (see equation 2); therefore, CBR discs fragment despite the fact that their spiral arms are weaker than in the discs in the CBF and CS models.

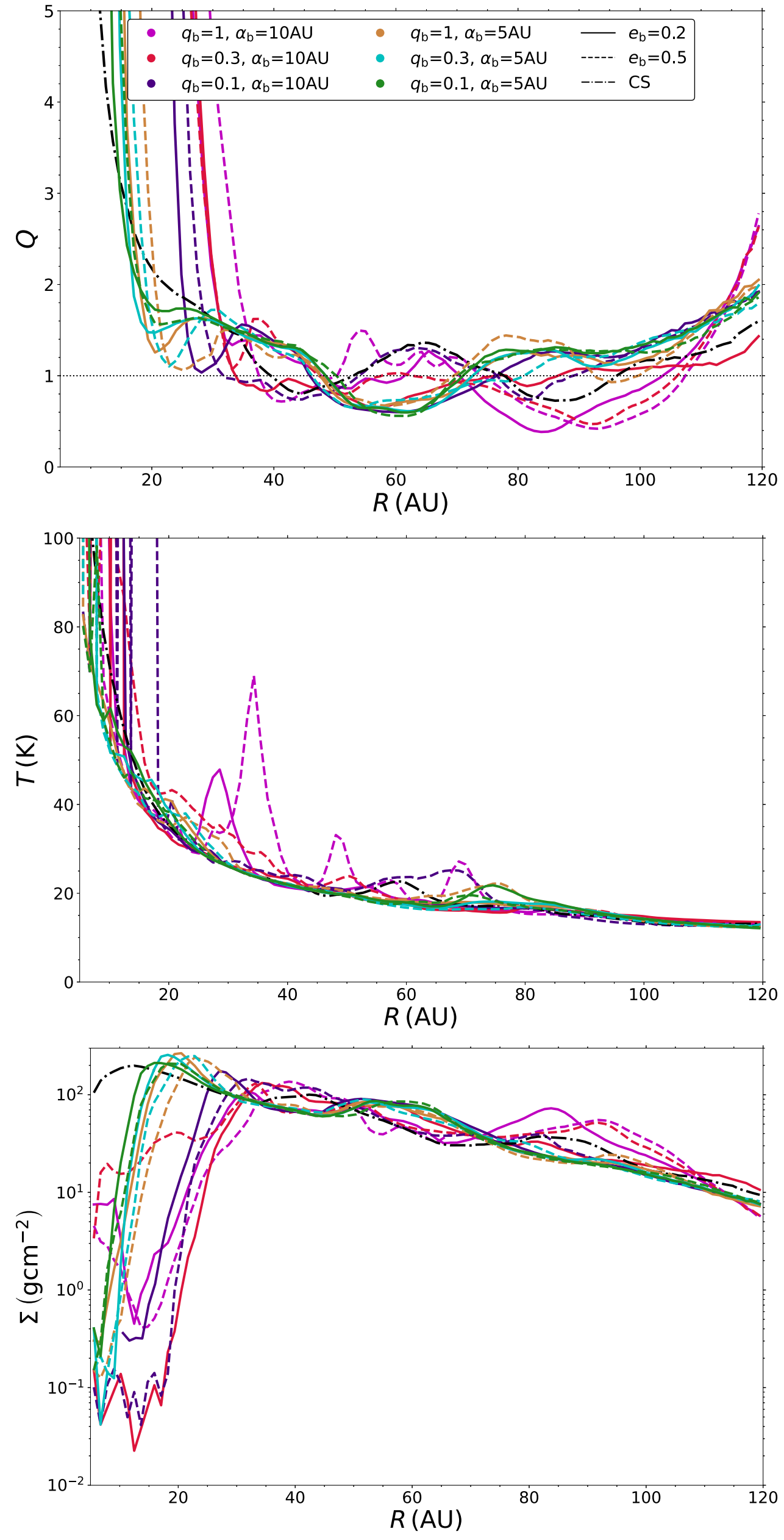

**Figure 7.** The Toomre parameter, temperature, and surface density of the discs in representative simulations for the CS and CBF models, plotted against the distance from the centre of mass of the binary/central star (surface density plots are shown in Figs 2, 3, and 4).

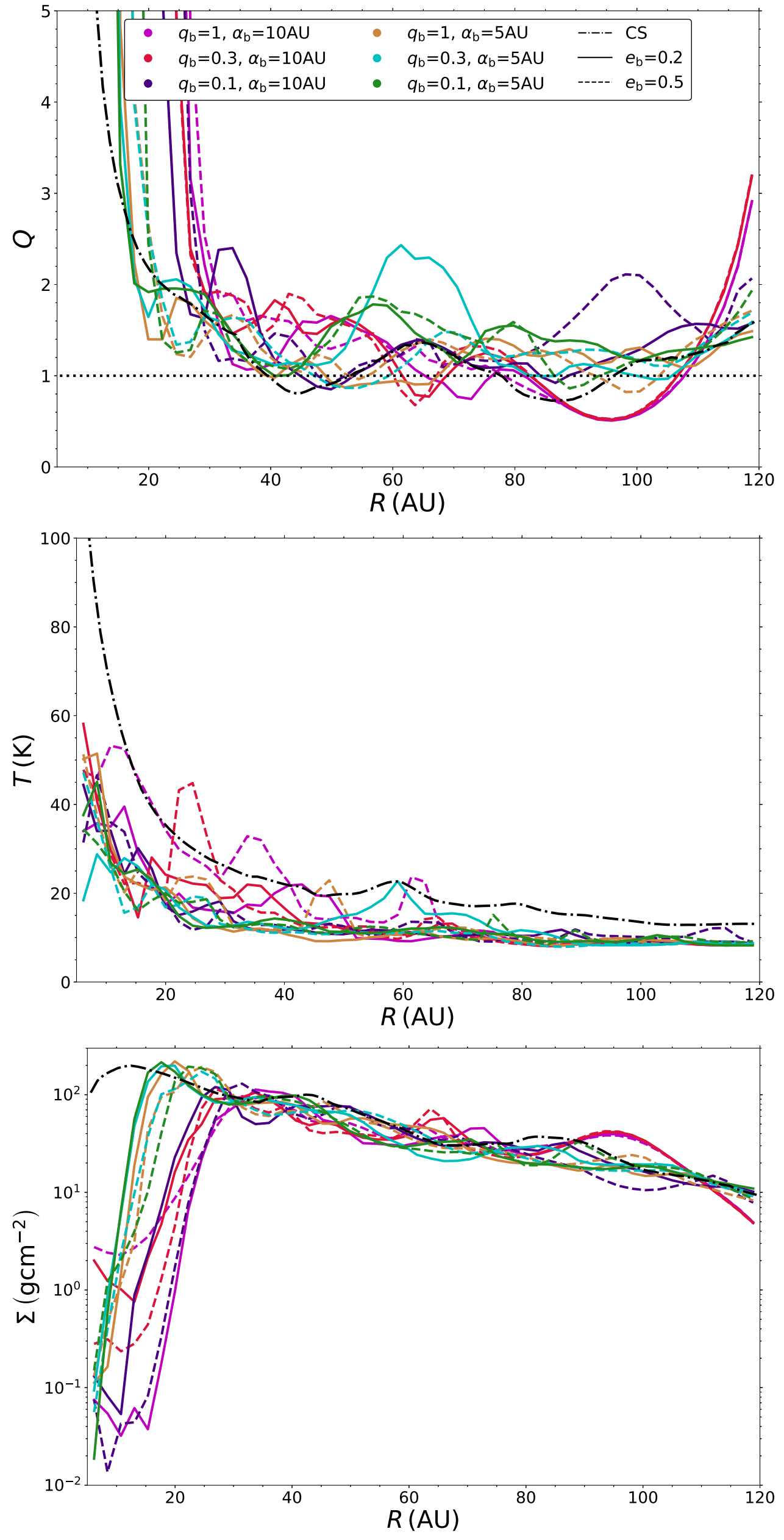

**Figure 8.** The Toomre parameter, temperature, and surface density of the discs in representative simulations for the CS and CBR models, plotted against the distance from the centre of mass of the binary/central star (surface density plots are shown in Figs 2, 5, and 6).

The regulation of disc fragmentation by gas cooling is also demonstrated by the fact that although discs around close binaries ($\alpha_b = 5$ au), have in general, spiral arms with larger peak amplitudes than discs around wider binaries ($\alpha_b = 10$ au) (which is consistent with M. Teasdale & D. Stamatellos 2023), the lower disc mass limit does not show any such dependence.

## 6 CONCLUSIONS

We used the SPH code SEREN to study the gravitational fragmentation of circumbinary discs. We performed three sets of simulations; the first covered circumstellar discs (circumstellar), the second covered circumbinary discs with the same temperature profile as the circumstellar discs (fiducial), and the third set covered circumbinary discs asymmetrically heated by each star of the binary individually (realistic). For each set of simulations, we varied the binary properties (separation, mass ratio, and eccentricity) to see their effect on the disc dynamics and fragmentation. The mass of the central object (star or binary) has been kept the same, 0.7 $M_\odot$, for all three models. Our aim is to investigate the lower mass limit for circumbinary disc

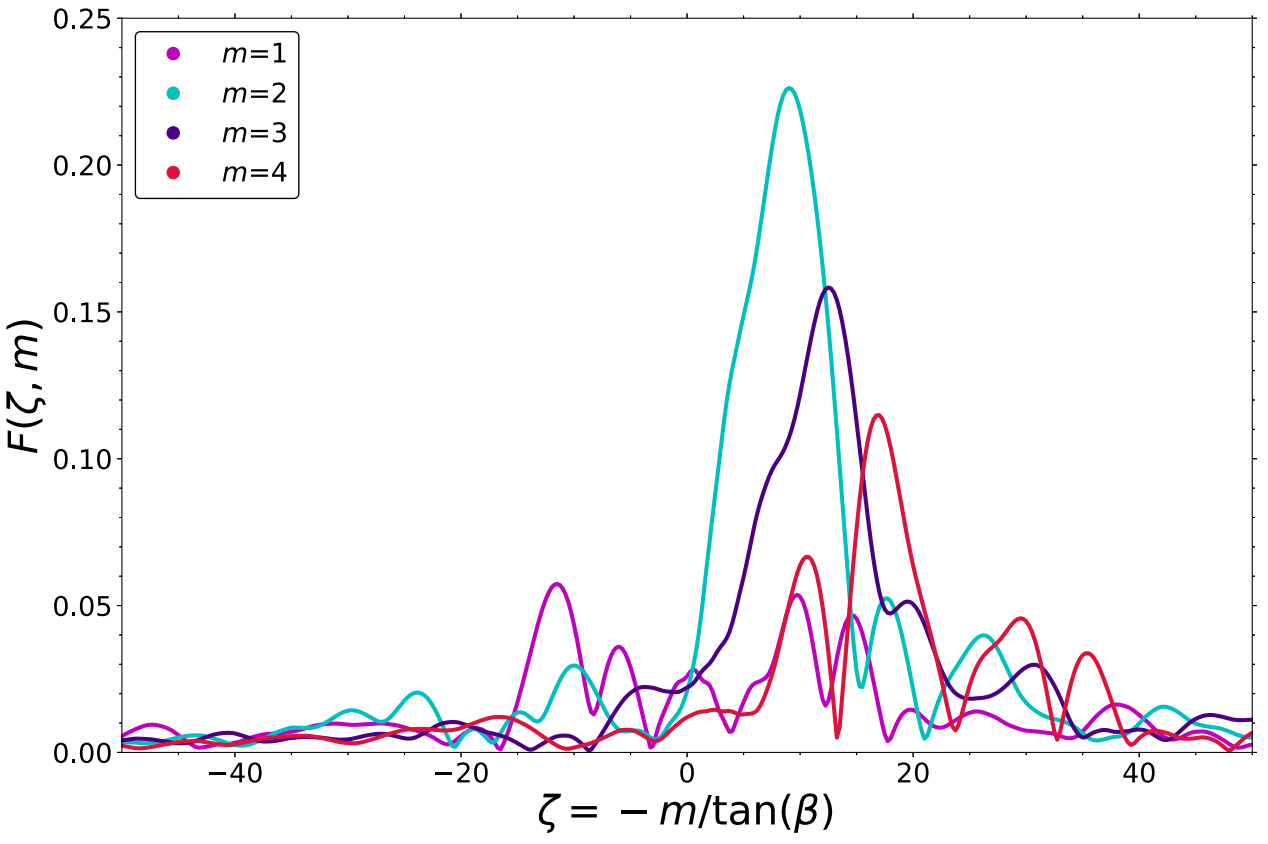

**Figure 9.** The strength of the different modes of spiral arms for a representative snapshot of the CBF model with a binary separation of $\alpha_{\rm b} = 10$ au, a binary mass ratio of $q_{\rm b} = 0.3$ and a binary eccentricity of $e_{\rm b} = 0.2$ (see Fig. 5 for the corresponding surface density plot). Here, the angle, $\zeta$, of the spiral is plotted against the amplitude, $F(\zeta, m)$, for all modes. The $m = 2$ mode is the dominant one.

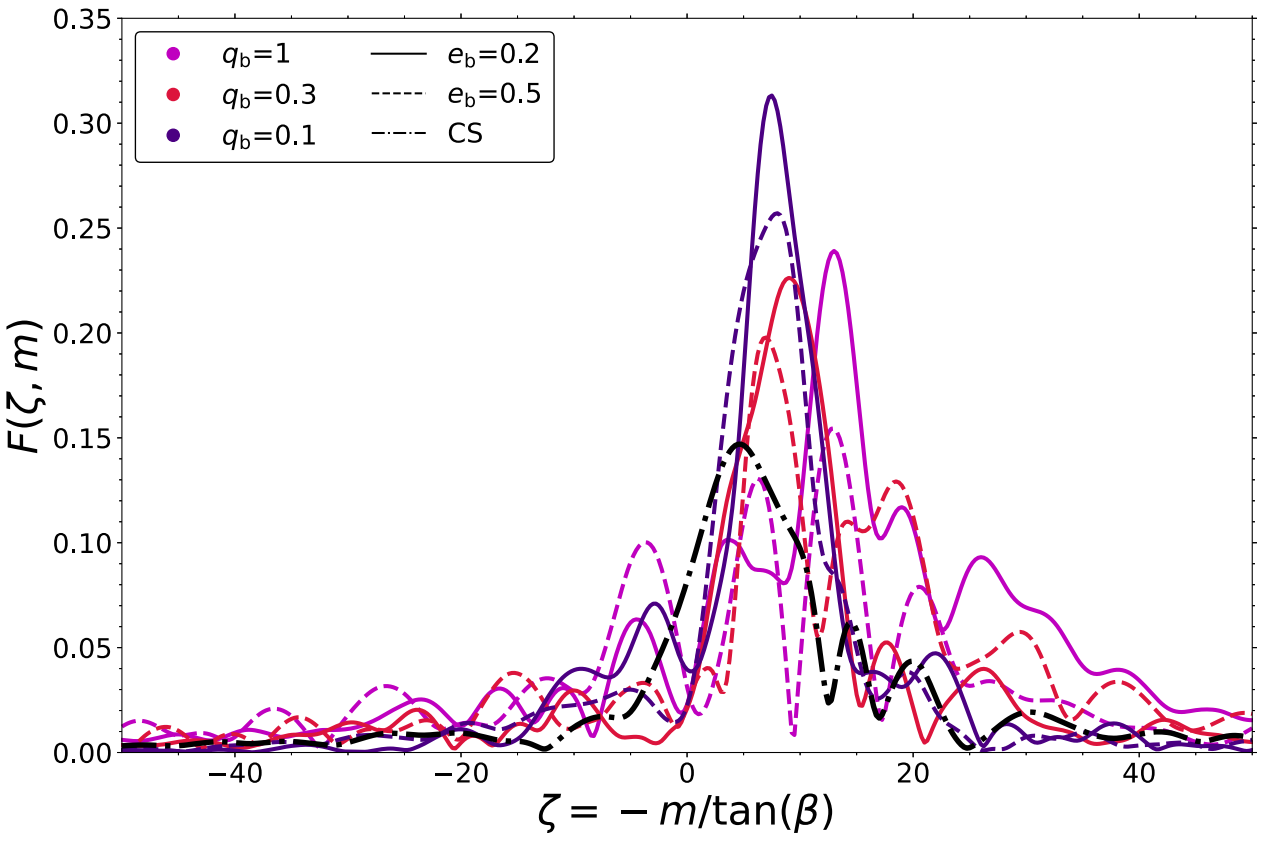

**Figure 10.** The strength of the $m = 2$ mode of the spiral arms for representative snapshots of the CBF model with a binary separation of $\alpha_{\rm b} = 10$ au (see Fig. 3 for the corresponding surface density plots). Here, the angle, $\zeta$, of the spiral is plotted against the amplitude, $F(\zeta, m)$, of the $m = 2$ mode. We also plot the spiral arm amplitude for one snapshot from the CS model for reference.

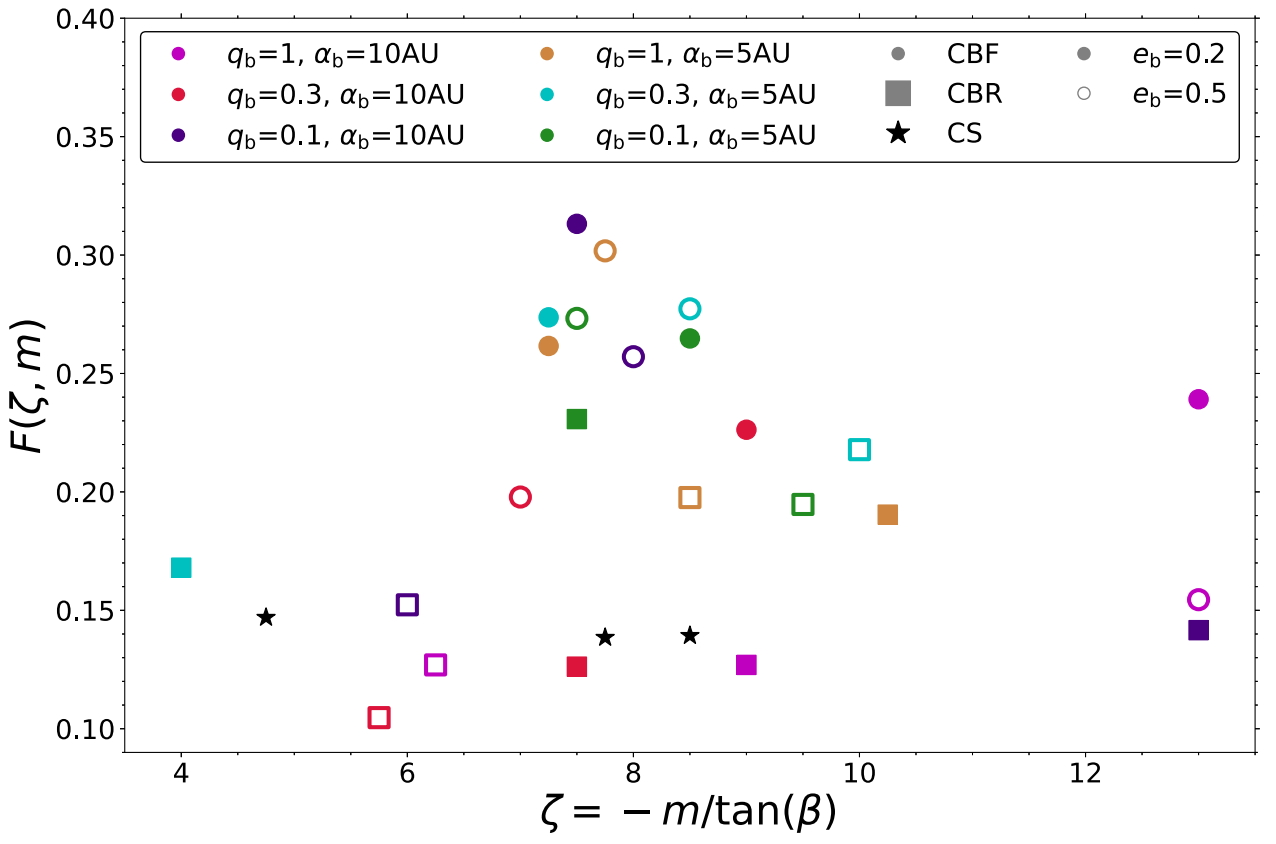

**Figure 11.** The peak strength of spiral arms for the circumstellar, circumbinary fiducial and realistic model simulations that are shown in Figs 2, 3, 4, 5, and 6. Here, the angle ($\zeta$) of the spiral is plotted against the maximum amplitude ($F(\zeta, m)$) of the $m = 2$ mode. There is no correlation between the spiral arm strength and the minimum disc fragmentation mass (e.g. CBF model discs show larger spiral arm strengths than the CS model discs but the minimum disc fragmentation mass is similar). This suggests that fragmentation is regulated by the disc cooling rather than the strength of the spiral arms (see discussion in the text).

fragmentation and compare it to the lower mass limit for circumstellar disc fragmentation.

We find that circumstellar discs fragment down to a disc-to-star mass ratio of $q_{\rm D} = 0.31$, which is in general agreement with previous studies (D. Stamatellos et al. 2011; J. Cadman et al. 2020; T. J. Haworth et al. 2020; A. Mercer & D. Stamatellos 2020). Similarly, circumbinary fiducial model discs are able to fragment down to $q_{\rm D} = 0.31$. On the other hand, realistic circumbinary discs fragment at a lower mass limit (by 45 per cent), at a disc-to-star mass ratio of 0.17–0.26, depending on the binary properties; a larger binary separation, mass ratio, and eccentricity promote fragmentation down to mass ratio of $q_{\rm D} = 0.17$. The lower disc mass limit for fragmentation is expected due to the lower disc temperature of the realistic CB models.

Furthermore, we find that fragmentation is regulated by cooling rather than the strength of the gravitational instability, as there is no correlation between the amplitude of the spiral arms and the lower disc masses needed for fragmentation. Indeed, fragmenting CBR discs show a lower spiral arm amplitude than fragmenting CBF and CS discs.

An example of a circumbinary disc that undergoes fragmentation is L1448 IRS3B (J. J. Tobin et al. 2016). This system consists of a binary with total mass of 1.19 $M_\odot$, attended by a 400 au disc. Evident spiral structure and a third object, IRS3B-c, embedded on one of the spirals, strongly suggests an *in situ* formation through disc instability. N. K. Reynolds et al. (2021) calculates the mass of the disc to be $\sim$0.29 $M_\odot$, so that the disc is gravitationally unstable at $>$120 au. The disc-to-star mass ratio of this system ($q_{\rm D} \sim 0.24$) is above the limit for fragmentation found by our models (both CBF and CBR), so our work supports the disc fragmentation scenario in this case.

We conclude that circumbinary discs fragment at lower disc masses than circumstellar discs, supporting the idea that some circumbinary planets may form by disc fragmentation. In a follow up paper, we will present the properties of the planets formed by fragmentation of circumbinary discs.

## ACKNOWLEDGEMENTS

We thank the anonymous referee for their constructive review that helped improve the paper. The simulations were performed using the University of Lancashire High Performance Computing (HPC) and High Throughput Computing (HTC) facilities. DS acknowledges support from STFC grant ST/Y002741/1. We thank David Hubber for the development of SEREN. Surface density plots were produced using SPLASH (D. J. Price 2007). This research has made use of data obtained from or tools provided by the portal exoplanet.eu of The Extrasolar Planets Encyclopaedia. This research has made use of the NASA Exoplanet Archive, which is operated by the California Institute of Technology, under contract with the National Aeronautics and Space Administration under the Exoplanet Exploration Program. This work was partially performed using resources provided by the Cambridge Service for Data Driven Discovery (CSD3) operated by the University of Cambridge Research Computing Service

(www.csd3.cam.ac.uk), provided by Dell EMC and Intel using Tier-2 funding from the Engineering and Physical Sciences Research Council (capital grant EP/T022159/1), and DiRAC funding from the Science and Technology Facilities Council (www.dirac.ac.uk).

## DATA AVAILABILITY

The simulation data used for this paper can be provided by contacting the authors.

This paper has been typeset from a TEX/LATEX file prepared by the author.